\documentclass[12pt,a4paper]{article}

\usepackage{amsmath}
\usepackage{cite}
\usepackage{graphicx}
\usepackage{hyperref}
\usepackage[hmargin=3cm,vmargin=3.5cm]{geometry}
\newcommand{\ket}[1]{\left\vert{#1}\right\rangle}
\newcommand{\braket}[1]{\left\langle{#1}\right\rangle}

\hypersetup{%
   pdftitle={Modern Feynman Diagrammatic One-Loop Calculations},
   pdfauthor={Thomas Reiter (speaker) for the Golem and the Samurai
              collaboration},
   pdfsubject={Proceedings of the ACPP meeting, Sep. 2010,
               KEK, Tsukuba, Japan},
   pdfkeywords={NLO, QCD, Golem, Samurai, Spinney, Haggies, LHC, Automatisation}
}

\title{Modern Feynman Diagrammatic One-Loop Calculations}
\author{\parbox{\textwidth}{
\begin{center}
Thomas Reiter$^1$,
G.~Cullen$^2$,
N.~Greiner$^3$,
A.~Guffanti$^4$,
J.P.~Guillet$^5$,
G.~Heinrich$^6$,
S.~Karg$^7$,
N.~Kauer$^8$,
T.~Kleinschmidt$^6$,
M.~Koch-Janusz$^1$,
G.~Luisoni$^6$,
P.~Mastrolia$^{9,10,11}$,
G.~Ossola$^{12}$,
E.~Pilon$^5$,
M.~Rodgers$^6$,
F.~Tramontano$^{11}$,
I.~Wigmore$^2$
\end{center}}\\[1.5ex]
\parbox{\textwidth}{\small
\begin{flushleft}
$^1$ Nikhef, 1098~XG Amsterdam, The Netherlands,\\
$^2$ School of Physics and Astronomy, %
	The University of Edinburgh,
	Edinburgh EH9\,3JZ, UK,\\
$^3$ Department of Physics, %
	University of Illinois at Urbana-Champaign, %
	Urbana IL, 61801, USA,\\
$^4$ Physikalisches Institut, %
	Albert-Ludwigs-Universit\"at, %
	79104~Freiburg, Germany,\\
$^5$ LAPTH, %
	74941 Annecy le Vieux Cedex, France,\\
$^6$ IPPP, %
	University of Durham, %
	Durham DH1\,3LE, UK,\\
$^7$ Institut f\"ur Theoretische Teilchenphysik und Kosmologie, %
	RWTH Aachen University, 52056~Aachen, Germany,\\
$^8$ Department of Physics, %
	Royal Holloway, University of London, %
	Egham TW20\,0EX, UK, \\
$^9$ Centro Studi e Ricerche ``E. Fermi'', 00184~Rome, Italy, \\
$^{10}$ Dipartimento di Fisica, Universit`a di Salerno,
	84084~Fisciano~(SA), Italy,\\
$^{11}$ Theory Group, Physics Department, CERN, 1211~Geneva 23, Switzerland,\\
$^{12}$ Physics Department, New York City College of Technology,
	City University Of New York, Brooklyn NY 11201, USA \\
\end{flushleft}
} % end parbox
}

\begin{document}
\maketitle
\begin{abstract}
In this talk we present techniques for calculating
one-loop amplitudes for multi-leg processes using Feynman diagrammatic
methods in a semi-algebraic context.
Our approach combines the advantages of the different methods
allowing for a fast evaluation of the amplitude while monitoring the
numerical stability of the calculation. In phase space regions close to
singular kinematics we use a method avoiding spurious
Gram determinants in the calculation.
As an application of our approach we report on the status of the
calculation of the amplitude for the process
$pp\rightarrow b\bar{b}b\bar{b}+X$.
\end{abstract}

\section{Introduction}
The experimental precision expected to be reached at the LHC demands
cross-section
calculations to move from the leading order approximation~(LO) to the
next-to-leading order~(NLO) for many processes with three, four or even
more particles in the final state.
The complexity introduced by both the real and virtual corrections
can only be overcome using automatised tools in the construction of
programs for the numerical evaluation of their matrix elements.

An important step towards the automatisation of NLO calculations was
the formulation of local subtraction terms which render real and virtual
corrections independently infrared finite. The two most commonly used
subtractions at the one-loop level are
the dipole formalism~\cite{Catani:1996vz,Catani:2002hc} and the
FKS subtraction method~\cite{Frixione:1995ms,Frixione:1997np}, both of which
have led to several packages for their automated implementation~\cite{%
Gleisberg:2007md,Seymour:2008mu,Hasegawa:2009tx,%
Frederix:2008hu,Frederix:2009yq,Frederix:2010cj,Gehrmann:2010ry,%
Czakon:2009ss}.
As one of their advantages, subtraction methods offer a well-defined way of
separating the different parts of a one-loop calculation. Inspired by this
separation, at the Les Houches workshop a standard interface between
Monte Carlo tools, computing the Born level, the real corrections and
the subtraction terms, and one-loop programs which only need to provide the
virtual corrections has been proposed~\cite{Binoth:2010xt}.

The persistent efforts spent on the improvement of the methods for the
calculation of one-loop amplitudes have recently induced a rapid growth in
the number of NLO predictions for the LHC. The most recent results with
four and more final state particles at the LHC are the corrections to
$pp\rightarrow b\bar{b}t\bar{t}+X$%
~\cite{Bredenstein:2008zb,Bredenstein:2009aj,Bevilacqua:2009zn},
$pp\rightarrow V+3j$%
~\cite{KeithEllis:2009bu,Berger:2009ep, Berger:2010vm},
$pp\rightarrow W^\pm+4j$%
~\cite{Berger:2010zx},
$pp\rightarrow W^+W^+jj$~\cite{Melia:2010bm}
and the $q\bar{q}$ initiated channel of
$pp\rightarrow b\bar{b}b\bar{b}+X$~\cite{Binoth:2009rv}.
However, many Standard Model processes of similar complexity which are relevant
for the discovery and/or the measurement of the properties of the Higgs boson
or new particles from extensions beyond the Standard Model~(BSM) are not yet
available at NLO~\cite{Binoth:2010ra}. Moreover, once new particles are found
at the LHC, their detailed study will require NLO studies in BSM models.
We have therefore developed and implemented methods for the automated
evaluation of one-loop Feynman diagrams, which we describe in
Section~\ref{sec:method}. In Section~\ref{sec:applications},
we present applications of our method, one of which is the calculation
of the gluon induced channel of the process
$pp\rightarrow b\bar{b}b\bar{b}+X$.

\section{Method}
\label{sec:method}

\subsection{A Numerator Representation for Feynman Diagrams}
Our method for the calculation of one-loop matrix elements is based on the
evaluation of Feynman diagrams. A one-loop diagram can be written as
\begin{equation}
\mathcal{D}=\int\!\!\frac{\mathrm{d}^nq}{i\pi^{n/2}}\frac{%
\mathcal{N}(q,\varepsilon)}{D_1\cdots D_N},
\end{equation}
where $D_i=[(q+r_i)^2-m_i^2+i\delta]$ and the integration momentum can be
split into its projection $\hat{q}^\mu$ on the
physical Minkowski space and an orthogonal component~$\tilde{q}$.
Instead of a single diagram, $\mathcal{D}$ can also refer to a group
of diagrams sharing the same set of denominators.
All quantities are defined such that the following equations hold:
\begin{subequations}
\begin{align}
g^{\mu\nu}=\hat{g}^{\mu\nu}+\tilde{g}^{\mu\nu},\quad
\hat{g}^\mu_\mu=4,\quad
\tilde{g}^\mu_\nu=-2\varepsilon,\\
\hat{g}^\mu_\rho\hat{g}^{\rho\nu}=\hat{g}^{\mu\nu},\quad
\tilde{g}^\mu_\rho\tilde{g}^{\rho\nu}=\tilde{g}^{\mu\nu},\quad
\hat{g}^\mu_\rho\tilde{g}^{\rho\nu}=0, \\
\hat{q}^\mu=\hat{g}^\mu_\nu q^\nu,\quad
\tilde{q}^\mu=\tilde{g}^\mu_\nu q^\nu,\quad
q^2=\hat{q}^2-\mu^2.
\end{align}
\end{subequations}
The numerator of the integral can then be expressed in terms of the quantities
$(\hat{q},\mu^2)$ as
\begin{equation}
\mathcal{N}(q,\varepsilon)=\mathcal{N}_0(\hat{q},\mu^2)
+\varepsilon\mathcal{N}_1(\hat{q},\mu^2)
+\varepsilon^2\mathcal{N}_2(\hat{q},\mu^2)
+\mathcal{O}(\varepsilon^3).
\end{equation}
The functions $\mathcal{N}_1$ and $\mathcal{N}_2$ are relevant in
regularisation schemes, where internal gauge bosons are kept in
$n$ dimensions ($n-2$ polarisations), such as the 't~Hooft-Veltman scheme
or na\"ive dimensional regularisation. The terms originating from
$\mathcal{N}_2$ are of purely infrared origin and therefore have to cancel
out, as shown in~\cite{Bredenstein:2008zb}. These terms can serve as an
additional check on the amplitude or simply be skipped in the computation.
In the following discussion there is no need to distinguish between the
individual functions $\mathcal{N}_i(\hat{q},\mu^2)$, as we can simply consider
the calculation of a diagram as separate calculations of the three terms in
$\mathcal{D}=\mathcal{D}_0+\varepsilon\mathcal{D}_1
+\varepsilon^2\mathcal{D}_2$. We refer to these functions collectively
as~$\mathcal{N}(\hat{q},\mu^2)$.

For the algebraic construction of $\mathcal{N}(\hat{q},\mu^2)$ from
the underlying Feynman rules we use QGraf~\cite{Nogueira:1991ex} for
generating the diagrams and Form~\cite{Vermaseren:2000nd} for their algebraic
simplification. Since we work with helicity projections of amplitudes,
we have devised a Form library for the manipulation of expressions
containing helicity spinors~\cite{Cullen:2010jv} which is publicly
available\footnote{\url{http://sourceforge.net/projects/spinney-form/}}.

In amplitudes containing external partons, the matrix element and therefore
each diagram is a vector in colour space
$\mathcal{D}=\sum_c\mathcal{D}^{c}\ket{c}$. We avoid working with the
vector components $\mathcal{D}^{c}$ by contracting each diagram with the
tree level amplitude, and working with the corresponding numerator functions
\begin{equation}
\mathcal{N}_i^{(\text{QCD})}(\hat{q},\mu^2)=
\sum_{c,c^\prime}\left(\mathcal{A}_\text{Born}^{c^\prime}\right)^\dagger
   \mathcal{N}_i^{c}(\hat{q},\mu^2)\braket{c^\prime\vert c}.
\end{equation}
This step is important to avoid a proliferation of numerical tensor reductions.
In fact, one cannot avoid working with the components $\mathcal{D}^{c}$ for
processes where the Born level vanishes and at the first non-vanishing order
one has to calculate the square of the NLO matrix element\footnote{%
We keep the name \emph{NLO} for the one-loop order even if the Born level
vanishes and the one-loop matrix element formally becomes the LO.}.

\subsection{Tensor Reduction at the Integrand Level}
\label{ssec:samurai}
Complete reducibility of one-loop diagrams to scalar
integrals implies
that the $q$-dependence of the numerator can be expressed in terms of
the denominators $D_i$~\cite{Ossola:2006us,Ossola:2007bb,Ellis:2007br,
Giele:2008ve,Melnikov:2010iu}:
\begin{multline}\label{eq:decompn}
\mathcal{N}(\hat{q},\mu^2)=
\sum_{j_1<\ldots<j_{N-1}}a_{j_1\ldots j_{N-1}}(q)D_{j_1}\cdots D_{j_{N-1}}\\
+\sum_{j_1<\ldots<j_{N-2}}b_{j_1\ldots j_{N-2}}(q)D_{j_1}\cdots D_{j_{N-2}}\\
+\sum_{j_1<\ldots<j_{N-3}}c_{j_1\ldots j_{N-3}}(q)D_{j_1}\cdots D_{j_{N-3}}\\
+\sum_{j_1<\ldots<j_{N-4}}d_{j_1\ldots j_{N-4}}(q)D_{j_1}\cdots D_{j_{N-4}}\\
+\sum_{j_1<\ldots<j_{N-5}}d_{j_1\ldots j_{N-5}}(q)D_{j_1}\cdots D_{j_{N-5}}.
\end{multline}
The calculation of a one-loop diagram therefore amounts to the extraction of
the coefficients, which are the residues of the integrand at the cuts
$D_{j_1}=\ldots=D_{j_{r}}=0$. A suitable parametrisation of the coefficients
and a fast implementation of the method in terms of a
\texttt{Fortran\,90}
library \textsc{Samurai}\footnote{\url{https://samurai.web.cern.ch/samurai/}} 
is described in~\cite{Mastrolia:2010nb}.

The decomposition given in Equation~\eqref{eq:decompn} works with
$n$-dimensional rather than four-dimensional propagators
$\hat{D}_i=[(\hat{q}+r_i)^2-m_i^2]$ and keeps the dependence on $\mu^2$
in the numerator. This improvement over the original OPP
method~\cite{Ossola:2006us,Ossola:2007bb} allows the determination of
the amplitude including both the cut-constructible and the rational part.
The \textsc{Samurai} implementation also includes several reconstruction tests
which provide a measure for the quality of the numerical result for
a given diagram. This information can be used to dynamically switch to
an alternative reduction method when the precision of the reconstruction
is not sufficient. Typically, unitarity based reduction methods fail in the
neighbourhood of kinematical points with vanishing Gram determinants.
Traditionally, implementations of unitarity methods go to higher precision
when such a case is detected. The following two sections introduce a method
which avoids the use of higher precision implementations.

\subsection{Improved Algebraic/Numerical Tensor Reduction}
The shortcomings of tensor reduction methods in the neighbourhood of vanishing
Gram determinants can be cured either by
expanding tensor coefficients about limits of vanishing Gram
determinants~\cite{Denner:2005nn} or by extending the integral basis
by integrals which are non-scalar but which have fast and stable
implementations~\cite{Binoth:1999sp,Binoth:2005ff,Binoth:2006hk}.
The latter approach has been followed in implementing the \texttt{Fortran\,95}
library Golem95~\cite{Binoth:2008uq}\footnote{%
\url{http://lappth.in2p3.fr/Golem/golem95.html}}, which in its
current version can be used for the reduction of tensor integrals with
up to six external legs for massive and massless loop-propagators.
For some $R\leq N$, the numerator structure of a Feynman diagram in a gauge
theory can always be written as
\begin{equation}\label{eq:tensgolem}
\mathcal{D}_i=\int\!\!\frac{\mathrm{d}^nq}{i\pi^{n/2}}\frac{%
\sum_{r=0}^RC^{(r)}_{\mu_1\ldots\mu_r}q^{\mu_1}\cdots q^{\mu_r}}{%
D_1\cdots D_N}=
\sum_{r=0}^RC^{(r)}_{\mu_1\ldots\mu_r}I_N^{n,\mu_1\ldots\mu_r},
\end{equation}
where the tensor integrals have the decomposition
\begin{multline}
I_N^{n,\mu_1\ldots\mu_r}=
\sum_{j_1\ldots j_r}\left[r_{j_1}\cdots r_{j_r}\right]^{{\mu_1\ldots\mu_r}}
   A^{N,r}_{j_1\ldots j_r}
+\sum_{j_1\ldots j_{r-2}}\left[g^{\cdot\cdot}
   r_{j_1}\cdots r_{j_{r-2}}\right]^{{\mu_1\ldots\mu_r}}
   B^{N,r}_{j_1\ldots j_{r-2}}\\
+\sum_{j_1\ldots j_{r-4}}\left[g^{\cdot\cdot}g^{\cdot\cdot}
   r_{j_1}\cdots r_{j_{r-4}}\right]^{{\mu_1\ldots\mu_r}}
   C^{N,r}_{j_1\ldots j_{r-4}}.
\end{multline}
The coefficient functions $A^{N,r}$, $B^{N,r}$ and $C^{N,r}$ are
implemented in the Golem95 library, which apart from these tensor
coefficients, also provides all the scalar integrals,
both finite and divergent ones\footnote{regulated dimensionally}, involving
massless and massive particles with real masses.
An extension of the library which, amongst other improvements,
allows for complex masses will be released soon.

\subsection{Tensorial Reconstruction at the Integrand Level}
\begin{figure}[!hbt]
\begin{center}
\includegraphics[height=3cm]{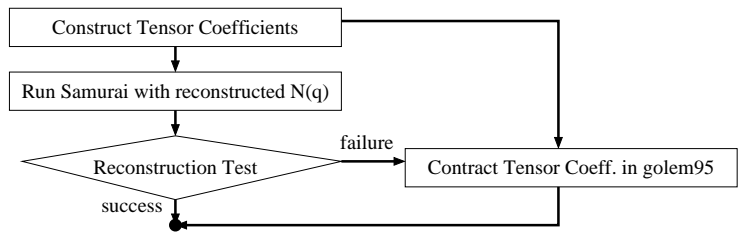}
\end{center}
\caption{Possible ways of using \textsc{Samurai} and Golem95 together
with tensorial reconstruction. On the left branch \textsc{Samurai} is
called with the numerator reconstructed according to Eq.~\eqref{eq:tensrec},
which can speed up the calculation. If the consistency check in
\textsc{Samurai} fails Golem95 is called as a rescue system. On the right
branch the call to \textsc{Samurai} is skipped and Golem95 is used as the
sole tensor reduction method.}\label{fig:flowchart}
\end{figure}

In Equations \eqref{eq:decompn} and \eqref{eq:tensgolem} we introduced
two different decompositions of the numerator function
$\mathcal{N}(\hat{q},\mu^2)$, which can be rewritten~as
\begin{equation}\label{eq:tensrec}
\mathcal{N}(\hat{q},\mu^2)=\sum_{\alpha=0}^{\lfloor R/2\rfloor}
\left(\mu^2\right)^\alpha\sum_{r=0}^{R-2\alpha}
\hat{C}^{(r,\alpha)}_{\mu_1\ldots\mu_r}
\hat{q}^{\mu_1}\cdots\hat{q}^{\mu_r}
\end{equation}
It turns out that a method for the numerical reconstruction of the tensor
coefficients $\hat{C}^{(r,\alpha)}$ can be constructed by means of
extracting the coefficients of a multivariate polynomial in the variables
$(\hat{q}^0,\ldots,\hat{q}^3,\mu^2)$.
As explained in Section~\ref{ssec:samurai},
a numerical implementation of $\mathcal{N}(\hat{q},\mu^2)$ is sufficient
as input for such a reconstruction, which is described in detail
in~\cite{Heinrich:2010ax}. For nearly all relevant cases tensorial
reconstruction~\eqref{eq:tensrec} requires fewer samplings
of the numerator function than reduction at the integrand
level~\eqref{eq:decompn}. An improvement of the \textsc{Samurai} algorithm
can therefore be achieved by passing a numerator function to the tensor
reduction which has been obtained from the original numerator by
tensorial reconstruction. Another possibility is to combine \textsc{Samurai}
and Golem95 as shown in Figure~\ref{fig:flowchart}.

\section{Applications}
\label{sec:applications}
\subsection{Automated One-Loop Matrix Element Evaluation}
We have combined the components described in Section~\ref{sec:method} to
build an automated general one-loop evaluator for matrix elements~(GOLEM).
The implementation is realised in form of a \texttt{python} program called
\texttt{golem-2{.}0}~\cite{Binoth:2008gx,Reiter:2009kb,Reiter:2009dk}.
We have recently added an interface for importing
Feynman rules from FeynRules~\cite{Christensen:2008py}. Furthermore,
the program supports the Binoth Les Houches Accord~\cite{Binoth:2010xt}
which defines a standard way of communicating information between Monte
Carlo generators and one-loop matrix element codes.
The numerical code is generated in \texttt{Fortran\,95} using the optimising
code generator \texttt{haggies}~\cite{Reiter:2009ts} but can be linked
both by the \texttt{Fortran}~\cite{Fortran:2003} and \texttt{C}
conventions.

\subsection{Results for \texorpdfstring{$pp\rightarrow b\bar{b}b\bar{b}+X$}{%
pp to b+b-bar+b+b-bar}}
At the LHC the signature of four $b$-jets can be a significant background
for Higgs searches in models beyond the Standard
Model~\cite{Binoth:2010ra,Lafaye:2000ec,Krolikowski:2008qa}.
Owing to large uncertainties and huge backgrounds
this search channel has been neglected in most phenomenological studies
for the LHC. Since in some BSM scenarios one of the Higgs bosons can be
predominantly decay into a four-$b$ final state,
it will be crucial to include the NLO corrections in $\alpha_s$ in future
studies.

In our setup, the virtual matrix elements have been generated using
\texttt{golem-2{.}0}
for the processes $q\bar{q}/gg\rightarrow q_1\bar{q}_1q_2\bar{q}_2$, where
$q$, $q_1$ and $q_2$ represent different quark flavours. The amplitudes
for the $b\bar{b}b\bar{b}$ final state have been obtained using the
relation~\cite{Reiter:2009kb}
\begin{equation}
\mathcal{A}_{b\bar{b}b\bar{b}}(1,2;3,4,5,6)=
\mathcal{A}_{q_1\bar{q}_1q_2\bar{q}_2}(1,2;3,4,5,6)
-\mathcal{A}_{q_1\bar{q}_1q_2\bar{q}_2}(1,2;3,6,5,4),
\end{equation}
where a permutation of the momenta encodes a simultaneous permutation
of momenta, colour and helicity labels. This trick reduces the number of
diagrams and therefore the code size by a factor two. The Born level amplitude
and the real emission corrections were calculated using
\texttt{MadGraph}~\cite{Stelzer:1994ta} and
\texttt{MadEvent}~\cite{Maltoni:2002qb}; for the subtraction
of the infrared singularities
\texttt{MadDipole}~\cite{Frederix:2008hu,Frederix:2010cj} has been employed.
On top of the programs described in Section~\ref{sec:method}, the library
\texttt{OneLOop}~\cite{vanHameren:2010cp} was used for the evaluation of the
scalar integrals.
The virtual corrections were integrated over phase space by reweighting
sets of unweighted Born level events~$\{p_i\}$. In order to obtain the virtual
corrections for an observable defined by the measurement
function~$\mathcal{F}(\{p_i\})$ we evaluate the Monte Carlo sum
\begin{equation}
\langle\mathcal{F}\rangle_{\text{virt}}=
\frac{\sigma_{\text{Born}}}{N}
\sum_{i=1}^N\left(1+\frac{
\mathcal{A}^\dagger_{\text{Born}}\mathcal{A}_{\text{virt}}+h.c.
}{
\mathcal{A}^\dagger_{\text{Born}}\mathcal{A}_{\text{Born}}}\right)
\mathcal{F}(\{p_i\}).
\end{equation}
In order to obtain
an estimate for the computational cost and the numerical stability
of the generated code, we have evaluated the matrix element for
500 points, starting from a fixed kinematics as specified in~\cite{Gong:2008ww}
and rotating the final state particles about the $z$-axis which coincides
with the direction of the incoming particles.
\begin{figure}
\begin{center}
\includegraphics[height=6cm]{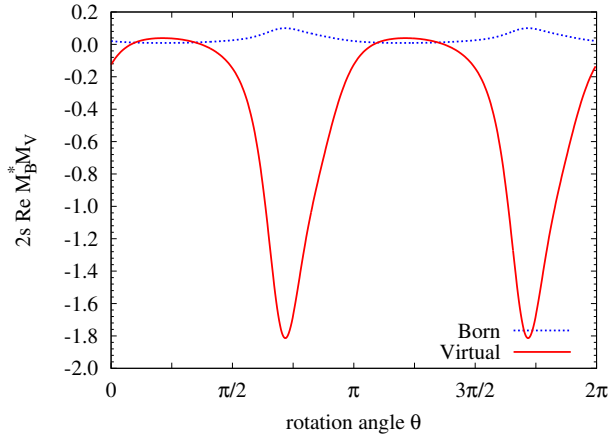}
\end{center}
\caption{Born (dashed) matrix element squared and the interference
term between Born and virtual corrections (straight line) of the
$gg\rightarrow b\bar{b}b\bar{b}$ amplitude. The curves have been obtained
by rotating the final state particles about the $z$-axis. The amplitude peaks
around $\theta_0\approx 2{.}32$ and $\theta_1=\theta_0+\pi$ which
are points close to double parton scattering kinematics.
}
\label{fig:plot}
\end{figure}

The results for the gluon induced subprocess are shown in Figure~\ref{fig:plot}.
As expected~\cite{Gong:2008ww,Bernicot:2007hs} the amplitude develops a
relatively sharp peak around $\theta_0\approx2{.}32$ and at $\theta_1=\theta_0+\pi$
because these points are close to a double parton scattering kinematics.
On a Xeon~5500 CPU the computation of a single phase space point in our setup
requires 18~seconds for the virtual corrections and 11\,ms for the
real corrections including subtractions.

\section{Conclusion}
In this talk we have reported on the status and various developments
in the Golem and the \textsc{Samurai} project. \textsc{Samurai} is a
program for the unitarity based reduction of one-loop amplitudes which,
amongst other improvements, preserves the $n$-dimensional information of
the numerator and is therefore capable of evaluating the amplitude
including all rational terms. The tensor reduction library Golem95 has
been extended with integrals with massive loop propagators; a new version
extending the integrals further to the case of complex masses will be
released soon. We have presented a
method for the numerical reconstruction of tensor coefficients from the
numerator of the integrand of a one-loop Feynman diagram, which can be used
to combine the advantages of unitarity based and conventional tensor
reduction methods. These different tensor reduction strategies are
combined in the one-loop matrix element generator \texttt{golem-2{.}0},
which has been used for the calculation of the QCD corrections
of the process $pp\rightarrow b\bar{b}b\bar{b}+X$.
Phenomenological results for this process are to be expected soon.

\bibliographystyle{hepshort}
\bibliography{acpp-reiter}

\end{document}